\documentclass[twocolumn,showpacs]{revtex4}

\usepackage{graphicx}%
\usepackage{dcolumn}
\usepackage{amsmath}

\makeatletter
\def\btt#1{\texttt{\@backslashchar#1}}%
\DeclareRobustCommand\bblash{\btt{\@backslashchar}}%
\makeatother

\begin{document}

\preprint{HEP/123-qed}

\title[Short Title]
{Partial and macroscopic phase coherences in underdoped Bi${}_{2}$Sr${}_{2}$CaCu${}_{2}$O${}_{8+{\delta}}$ thin film
}

\author{
H.~Murakami${}^{1}$,~T.~Kiwa${}^{1}$,~N.~Kida${}^{1}$,~M.~Tonouchi${}^{1,2}$,~T.~Uchiyama${}^{3}$,~I.~Iguchi${}^{3}$,~and Z. Wang${}^{4}$,
}
\affiliation{%
Research~Center~for~Superconductor~Photyonics,~Osaka~University${}^{1}$~and~CREST-JST${}^{2}$, Suita,~Osaka~565-0871,~Japan
Department~of~Physics,~Tokyo~Institute~of~Technology~and~CREST-JST${}^{3}$,~2-12-1~Ookayama,~Meguro-ku,~Tokyo~152-8551,~Japan 
Kansai~Advanced~Research~Center,~Communication~Research~Laboratory${}^{4}$,~Kobe~651-2492,~Japan
}%

\date{\today}

\begin{abstract}
A combined study with use of time-domain pump-probe spectroscopy and time-domain terahertz transmission spectroscopy have been carried out on an underdoped Bi$_2$Sr$_2$CaCu$_2$O$_{8+{\delta}}$ thin film. It was observed that the low energy multi-excitation states were decomposed into superconducting gap and pseudogap. The pseudogap locally opens below $T^*{\simeq}210$ K simultaneously with the appearance of the high-frequency partial pairs around 1.3 THz. With decreasing temperature, the number of the local domains with the partial phase coherence increased and saturated near 100 K, and the macroscopic superconductivity appeared below 76 K through the superconductivity fluctuation state below 100 K. These experimental results indicate that the pseudogap makes an important role for realization of the superconductivity as a precursor to switch from the partial to the macroscopic phase coherence.

\end{abstract}


\maketitle

Although a great deal of experimental and theoretical efforts have been made to shed light on the normal state anomalies caused in underdoped high-temperature superconductors (HTSC), it is still an open field of research~\cite{alloul,puchkov,uchida,renner,ding,demsar}. The main focus of the studies are the possibility of the pseudogap opening at a characteristic temperature ${T^*}$ well above the superconducting transition temperature ${T_C}$ and the relation to the macroscopic phase coherence realized below $T_C$. It is generally considered that the pseudogap accompanying by a substantial reduction in the density of state (DOS) is caused by partial pair formation or partial charge (or spin) ordering in the crossover regions from antiferromagnetic ordering to metallic state. 

To investigate such a complicate multi-excitation system where the superconducting gap ${\Delta}_s$ and the pseudogap ${\Delta}_p$ may coexist, time-domain pump-probe spectroscopy (TDPPS) is a very useful tool~\cite{kabanov,han,easley,gong,nashima}. It has a potential to decompose the multi-excitation processes with different relaxation dynamics for its excellent time resolution better than 0.1 ps, and to bring some important information on the low-lying electronic structure~\cite{kabanov}. 

On the other hand, it has been reported that such relaxation processes of excited carriers (or quasi-particles) as concerned with these gaps take place within a few picoseconds in the case of YBa$_2$Cu$_3$O$_{7-{\delta}}$ (YBCO)~\cite{demsar}. Therefore, some complementary information about the low-energy-excitation states may be obtained by investigating the high-frequency properties in terahertz (THz) frequency regions. It can be very effectively studied using time-domain THz transmission spectroscopy (TDTTS). 

In present study we performed TDPPS and TDTTS measurements on an underdoped Bi$_2$Sr$_2$CaCu$_2$O$_{8+{\delta}}$ (BSCCO) thin film prepared by pulsed laser deposition (PLD) method, in order to investigate the existence of pseudogap and its relation to the superconductivity. In this Letter we report the experimental evidences for coexistence of two excitation processes due to ${\Delta}_s$ and ${\Delta}_p$ below $T_C$ and for partial-phase coherence realized simultaneously with the pseudogap opening below ${T^*{\simeq}210}$ K. 

\begin{figure}
\includegraphics[width=75mm]{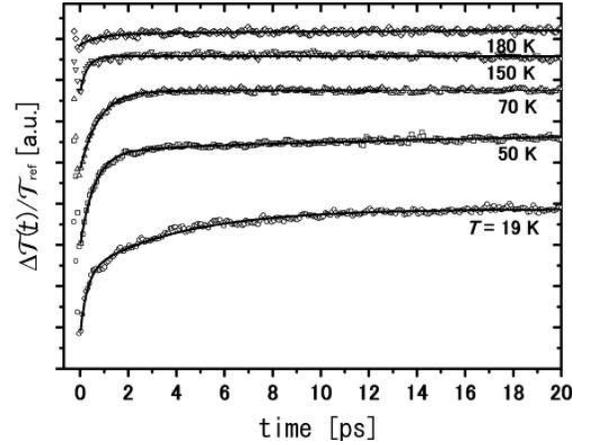}
\caption{Transient transmissivity changes as a function of delay time measured below and above $T_{C1}$. The lines show the fits using exponential decay functions with the relaxation times of ${\tau}_s$ ($T<T_{C1}$) and ${\tau}_p$ ($T<T^*$).}
\label{sp}
\end{figure}

For the measurement, $c$-axis oriented BSCCO thin film of 50${\pm}$4 nm-thick (PLD-BSCCO) was prepared on a MgO substrate by PLD method~\cite{uchiyama}. The thin film characterization was carried out by X-ray diffraction (XRD), critical supercurrent density (${J_C}$) measurement, and atomic force microscopy. XRD patterns showed that all diffraction peaks were indexed by (00${l}$) of BSCCO single phase. From the detail analysis, the lattice constant along the $c$-axis of thin film was determined as $c$=3.088 nm which was a little longer than $c$=3.08 nm for an optimally doped BSCCO crystal. The long lattice constant shows the characteristic peculiar to underdoped BSCCO. 

The temperature dependence of resistivity showed superconducting transition at ${T_{C1}=76}$ K and ${T_{C2}=58}$ K, corresponding to the current density of ${J_{C1}=33}$ A/cm${}^{2}$, and ${J_{C2}=4{\times}10^{3}}$ A/cm${}^{2}$, respectively. The current density gradually increased below ${T_{C1}}$ and then rapidly increased below ${T_{C2}}$ with decreasing temperature. The observed maximum value of ${J_{Cmax}=4.3{\times}10^5}$ A/cm${}^{2}$ at ${T}$ = 10 K was about one order of magnitude lower than that of optimally doped BSCCO film described in Ref.~[\onlinecite{otsuka}].

In the TDPPS measurement on HTSCs, the photo-excited carriers by pump pulse quickly relax to the upper excitation state of the low energy excitation gap, such as ${\Delta}_s$ and ${\Delta}_p$, near the Fermi level by electron-electron and electron-phonon scatterings. In these relaxation processes a lot of low-energy-excited carriers are created by an avalanche effect in the non-equilibrium superconductivity. Since this first relaxation stage breaks up within ${\sim}$0.1 ps~\cite{demsar, kabanov}, we can analyze in detail the following relaxation processes of the low-energy-excited carriers from the excitation states of respective gaps. These relaxation processes can be observed as the transient change in the transmissivity of probing pulse with the time-resolution of about 0.1 ps. In the experiment, pump and probe pulse beams were emitted from a mode-locked Ti:sapphire laser (Tsunami; Spectra-Physics) which produces optical pulses with width of about 100 fs and wavelength of 810 nm operating at a repetition rate of 82 MHz. To get time-resolved spectrum of transmissivity, the time-delay of probing-pulse was varied using a sliding reflection mirrors computer-controlled with a stepping motor. The polarization of the pumping laser beam was rotated by 90$^{\circ}$ with use of a ${\lambda}$/2 plate so that the polarizations for the pump and probe beams were set to be perpendicular each other. The pump and probe beams were focused at a same spot on the sample surface at about 60 ${\mu}$m and 40 ${\mu}$m in diameter with the averaged powers of about 0.36 nJ and 0.1 nJ per respective pulses.

To detect the time-resolved transmissivity, the intensity of the transmitting probe pulse, $\mathcal{T}$($t$), and the reference pulse, $\mathcal{T}_{ref}$, were detected by silicon PIN photo-diodes, and the difference, ${\Delta}\mathcal{T}(t)$=$\mathcal{T}(t)-\mathcal{T}_{ref}$, was lock-in detected by chopping the pumping pulses at 2 kHz. Further, $\mathcal{T}_{ref}$ was adjusted to the value of $\mathcal{T}(t)$ just before pump-pulse irradiation by using a ND filter. The details of the optical system have been reported before~\cite{nashima}.

In the TDTTS measurements, we used the same PLD-BSCCO thin film, and extracted the complex conductivity, ${\sigma}={\sigma}_1+i{\sigma}_2$, by numerically solving the complex transmittance without Kramers-Kronig analysis at frequencies from 0.3 to 1.5 THz and at temperatures from 4.2 K to room temperature. The details of our optical system and data processing have been reported elsewhere~\cite{kiwa}.

\begin{figure}
\includegraphics[width=75mm]{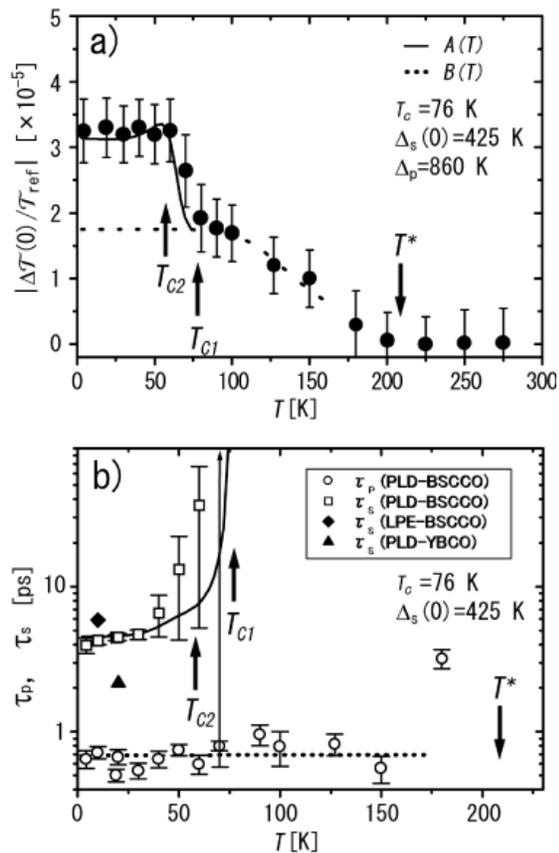}
\caption{
(a) Temperature dependence of the maximum amplitude, ${\mid}{\Delta}\mathcal{T}(0)/\mathcal{T}_{ref}{\mid}$. The solid [$A(T)$] and dotted [$B(T)$] lines show the fits using Eqs.~(2) and (3) in Ref.~[\onlinecite{demsar}]. Here, the solid line is shifted vertically for an appropriate value to compare with the experimental data. The values of ${\Delta}_s$(0) and ${\Delta}_p$ used in the fit are also shown.
(b) Temperature dependence of relaxation times, ${\tau}_p$ and ${\tau}_s$. The solid line shows the fit below $T_{C1}$ using Eq.~(1) in Ref.~[\onlinecite{demsar}] and the dotted line shows the mean value of ${\tau}_p$.
}
\label{tau}
\end{figure}

Typical transient transmissivity changes as a function of delay time, ${\Delta}\mathcal{T}(t)/\mathcal{T}_{ref}$, measured below and above $T_{C1}$ are plotted in Fig.~\ref{sp}. The sign of ${\Delta}\mathcal{T}(t)$ was negative on the contrary to the positive sign of ${\Delta}\mathcal{R}(t)$ in the reflectivity~\cite{murakamiISS}. We can see sudden decrease of ${\Delta}\mathcal{T}(t)$ at $t$=0 ps corresponding to the pump pulse irradiation. Figure~\ref{tau}(a) shows the plots of  the amplitude ${\mid}{\Delta}\mathcal{T}(0)/\mathcal{T}_{ref}{\mid}$ versus temperature. The temperature dependence of ${\mid}{\Delta}\mathcal{T}(0)/\mathcal{T}_{ref}{\mid}$ showed almost the same behavior with that of ${\Delta}\mathcal{R}/\mathcal{R}$ observed on Y$_{1-x}$Ca$_x$Ba$_2$Cu$_3$O$_{7-{\delta}}$ single crystals~\cite{demsar}. With decreasing temperature ${\mid}{\Delta}\mathcal{T}(0)/\mathcal{T}_{ref}{\mid}$ shows a gradual increase below $T^*{\sim}$210 K and a rapid increase near $T_{C1}$ after showing the tendency of saturation near 100 K, and then shows almost a constant value below $T_{C2}$. The observed behavior can be approximately explained by two excitation gaps using $T$-dependent superconducting gap ${\Delta}_s(T)$ and $T$-independent pseudogap ${\Delta}_p$ as reported by Demsar {\it et al.}~\cite{demsar}. They applied a two-component fit to the data with use of an exponential decay function, ${\Delta}\mathcal{R}(t)/\mathcal{R}=A(T){\exp}(-t/{\tau}_s)+B(T){\exp}(-t/{\tau}_p)$ ($A(T)=0$ for $T_{C1}<T$). Here, they considered temperature ($T$)-dependent ${\tau}_s$ and $T$-independent ${\tau}_p$ as the relaxation times of excited carriers of ${\Delta}_s$ and ${\Delta}_p$, respectively. The amplitudes, $A(T)$ and $B(T)$, are $T$-dependent and given by Eqs.~(2) and (3) in Ref.~[\onlinecite{demsar}], respectively, and these are relating to the magnitudes of the respective gaps and the number of the excited carriers. In Fig.~\ref{tau}(a) the fitting curves, $A(T)$ and $B(T)$, are also displayed. For the fits, we considered the gap parameters of ${\Delta}_s(0){\simeq}$37 meV and ${\Delta}_p{\simeq}$74 meV. These gap values are comparable with previous tunneling studies~\cite{renner, murakamiTUN, suzuki}. It is noted here that the superconductivity occurs followed by the temporary saturation in the amplitude near 100 K.

To investigate the detailed relaxation process from the data, the relaxation times were estimated by fit with use of the exponential decay function using $T$-dependent ${\tau}_s$ and ${\tau}_p$ by replacing ${\Delta}\mathcal{R}(t)/\mathcal{R}$ with ${\mid}{\Delta}\mathcal{T}(t)/\mathcal{T}_{ref}{\mid}$. Figure~\ref{sp} also shows the fitting curves. We could obtain a fairly good agreement between the data and the exponential decay functions over the wide temperature ranges utilized to present study. The mean value of ${\tau}_s$ was estimated as $4.2{\pm}0.3$ ps below 40 K, and that of ${\tau}_p$ as $0.7{\pm}0.3$ ps below 150 K. The estimated relaxation times are summarized in Fig.~\ref{tau}(b). The ${\tau}_s$ value diverged and disappeared near $T_{C1}$ with increasing temperature. On the other hand, ${\tau}_p$ also showed $T$-dependent characteristic. Namely, it was $T$-independent below 150 K, and diverged near $T^*$. The observed temperature dependence of ${\tau}_p$ indicates that the pseudogap is also $T$-independent below 150 K and closes near $T^*$. 

Summarizing the experimental results observed, we can conclude a few important points as follows. The number of local domains with coherent pair or charge ordering, which can produce the pseudogap structure in DOS, increases with decreasing temperature from $T^*$ to 100 K, because the amplitude ${\mid}{\Delta}\mathcal{T}(0)/\mathcal{T}_{ref}{\mid}$ in Fig.~\ref{tau}(a) substantially increases while ${\tau}_p$ keeps a constant value of about 0.7 ps. Furthermore, it is also recognized that the macroscopic phase coherence takes place over all the sample after the number of the local domains attains to a certain level. 

It is also important to compare the ${\tau}_s$ values with those of overdoped BSCCO and YBCO, because the relaxation time of quasi-particle is closely related to the bottleneck structure near the Fermi level~\cite{dynes}. These ${\tau}_s$ values observed in our previous studies are also plotted in Fig.~\ref{tau}(b)~\cite{murakamiISS}. Here, LPE-BSCCO was an overdoped BSCCO thin film grown by liquid phase epitaxy method.  In the fitting process for the data observed on several overdoped BSCCO samples, only longer-components corresponding to ${\tau}_s$ have been observed. The amplitudes corresponding to ${\mid}{\Delta}\mathcal{T}(0)/\mathcal{T}_{ref}{\mid}$ or ${\Delta}\mathcal{R}(0)/\mathcal{R}$ were also reduced to about 0 near $T_C$. As for the ${\tau}_s$ value, ${\tau}_s(10K)=5.9{\pm}0.2$ ps for LPE-BSCCO is longer than those of underdoped sample by about 1.7 ps. On the other hand, ${\tau}_s(19K)=2.1{\pm}0.1$ ps for PLD-YBCO shows a good consistency with those observed by Demsar {\it et al.}~\cite{demsar}, and it is about a half of those of PLD-BSCCO. It is generally considered that long relaxation time of quasi-particles yields a well-developed superconducting gap structure. In fact, the studies with use of a scanning tunneling microscope (STM) showed distinct superconducting gap structures for BSCCO single crystal reflecting the longer relaxation time ${\tau}_s$, while only smeared gaps for YBCO even on an extremely clean surface~\cite{renner, murakamiTUN}.

\begin{figure}
\includegraphics[width=75mm]{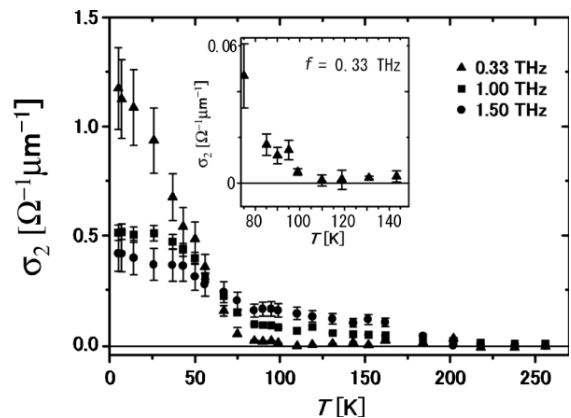}
\caption{
Temperature dependence of imaginary part of complex conductivity, ${\sigma_2}$, at several frequencies. The ${\sigma_2}$ shows a measurable increase below a characteristic temperature of ${{T^*}{\simeq}210}$ K for ${f}$= 1.0 and 1.5 THz, and a rapid increase below ${T_{C1}}$. The inset shows the detailed plots around 100 K for $f$=0.33 THz.
}
\label{sigma}
\end{figure}

On the other hand, the formation of pseudogap below $T^*$ was also supported by TDTTS measurement. Figure~\ref{sigma} shows the temperature dependence of the imaginary part of complex conductivity, ${\sigma_2}$, at several frequencies. It is noted here that at the higher-frequency, ${{\sigma}_{2}}$ ($f$=1.0 and 1.5 THz) shows a gradual increase below the corresponding temperature $T^*$ to pseudogap opening, and then a rapid increase below $T_{C1}$. On the other hand, at the lower-frequency, ${\sigma_2}$~(${f=0.33}$ THz) becomes observable below 100 K, and then rapidly increases below ${T_{C1}}$. Similar lower frequency behavior was observed by Corson {\it et al.}~\cite{corson}. They explained this phenomenon using the fluctuation in the phase correlation time reflecting the motion of thermally generated topological defects in the phase, or vortices. It is very interesting that the fluctuation takes place simultaneously with the saturation of the number of the pseudogap regions as shown in Fig.~\ref{tau}(a).

\begin{figure}
\includegraphics[width=75mm]{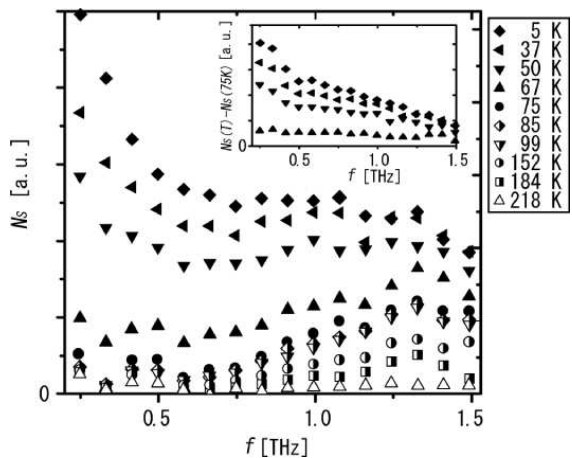}
\caption{
Frequency dependence of density of supercarrier, ${N_S}$, for several temperatures. Inset shows the density of supercarriers arising below 75 K.
}
\label{ns}
\end{figure}

For more detail investigation we simply estimated the density of supercarriers, ${N_S}$, by using the relationship of $N_S{\propto}{{\lambda_L}^{-2}}={\mu_0}{\omega}{\sigma_2}$. Here, ${\lambda_L}$ is the London penetration depth, ${\mu_0}$ is the magnetic permeability of vacuum space, and ${\omega}$ is the angular frequency. Figure~\ref{ns} shows the frequency dependence of ${N_S}$. Since we ignored the Drude contribution from the normal carriers to ${\sigma_2}$ for simplicity, the plotted data marked as blank (${T^*<T}$) may not have essential meaning. In the high frequency regions around a characteristic frequency of $f_C{\simeq}$1.3 THz, we can see a broad local maximum even at 75 K (${\simeq}T_{C1}$). It is noted that the $N_S$ ($f=f_C$) increases below $T^*$ and once saturates near 100 K (there is little difference between $N_S$(75K), $N_S$(85K) and $N_S$(99K)) with decreasing temperature. This behavior is also seen in the temperature dependence of ${\sigma}_2$ in Fig.~\ref{sigma}, and qualitatively consistent with the behavior of ${\mid}\mathcal{T}$(0)/$\mathcal{T}_{ref}{\mid}$ in Fig.~\ref{tau}(a). This means that the pseudogap opening is strongly concerned with the partial-coherent-ultrafast carriers around $f_C$. It is further interesting that the characteristic frequency $f_C$ just corresponds to the relaxation time of ${\tau}_p{\simeq}0.7$ ps. On the other hand, we can see which carriers are dominantly concerned with the macroscopic phase coherence below $T_{C1}$ by subtracting the $N_S$(75K) from $N_S(T)$ ($T<T_{C1}$). The inset shows the temperature and frequency dependence below $T_{C1}$. Here, we can see that the macroscopic phase coherence is the main contribution by lower frequency carriers. 

These obtained experimental results may allow us to conclude one possible scenario to realize the macroscopic superconductivity, as follows. The pseudogap opens below $T^*$ in the local domains with partial phase coherence closely related to the higher-frequency carriers around a characteristic frequency of $f_C{\simeq}$1.3 THz. With decreasing temperature the number of the domains increases and saturates near 100 K. In this condition the superconductivity fluctuation may take place by weak links of the order parameters of the local domains, and the macroscopic phase coherence realize over all the sample below $T_{C1}$ with the help of the lower-frequency carriers. To check this scenario, further complementary study with use of STM in the wide temperature ranges and on the wide surface area over several thousands ${\mu}$m${}^2$ may be useful. 

In summary, TDPPS and TDTTS studies were carried out on an underdoped BSCCO thin film. TDPPS decomposed the multi-excitation states into the superconducting gap and pseudogap, and TDTTS brought the complementary information about the carriers deeply concerned with the partial and the macroscopic phase coherence. These obtained experimental results show that the pseudogap formation makes an important role for realization of the macroscopic superconductivity as precursor to switch from the partial to the macroscopic phase coherence.

\begin{acknowledgments}
This work was supported in part by a Grant-in-Aid for Scientific Research (B), No.
12450146, from the Ministry of Education, Culture, Sports, Science and Technology. One of the authors (H.M.) wishes to acknowledge the support of The Murata Science Foundation.

\end{acknowledgments}

\bibliography{apssamp}

\end{document}